\documentclass[runningheads]{llncs}

\usepackage{graphicx}
\usepackage{amsmath}
\usepackage{amssymb}
\usepackage{booktabs}
\usepackage{xcolor}

\graphicspath{{./}{metFigs/}{resFigs/}{appFigs/}{figures/}}

\begin{document}

\title{A Shared IPTC Topic Space for Comparable Topic Modelling of Social and Mainstream Media}

\titlerunning{A Shared IPTC Topic Space for Cross-Source Topic Modelling}

\author{Din Iskakov\inst{1} \and
Sebastian Gon\c{c}alves\inst{2} \and
Marco Idiat\inst{2} \and
Mendeli Vainstein\inst{2} \and\\
Aline Villavicencio\inst{1} \and
Ronaldo Menezes\inst{1}$^{\star}$ \and
Rodrigo Wilkens\inst{1}}
\authorrunning{D. Iskakov et al.}
\institute{Department of Computer Science, University of Exeter, UK\\
\email{r.menezes@exeter.ac.uk, r.wilkens@exeter.ac.uk}
\and
Department of Physics, Federal University of Rio Grande do Sul, Brazil}

\maketitle

\renewcommand{\thefootnote}{$\star$}
\footnotetext{Corresponding authors.}
\renewcommand{\thefootnote}{\arabic{footnote}}

\begin{abstract}
Comparing topic attention across different media is hindered by a fundamental modelling problem: topic models fitted separately to each corpus produce corpus-specific topic spaces that cannot be aligned directly. This paper presents a reproducible framework that places corpora in a single shared topic space defined by a taxonomy. Discovered topics are obtained with guided BERTopic, scored against the ninety-four IPTC Media Topics' taxonomy topics (level-1) through weighted keyword and target centroids, and then collapsed upward to seventeen IPTC parent topics by a maximum-similarity rule. The framework was developed and selected on a controlled New York Times 2011 corpus through a narrowing sequence: a broad model screen, a focused mapping refinement, a strict finalist comparison, a target-construction ablation, and a threshold calibration. In this corpus, the guided family retained substantially stronger mapped coverage than a zero-shot benchmark under stricter assignment thresholds, a parent-enriched target construction improved both coverage and parent consistency, and coverage declined gradually rather than collapsing as the assignment threshold was tightened. The contribution is an externally anchored method for constructing a shared topic space that enables reproducible cross-source topic comparison. 
\end{abstract}

\section{Introduction}

Social media platforms such as Twitter and mainstream news organisations represent structurally different information flows. One is decentralised and bottom-up, the other editorially structured and top-down, and although they often address the same subjects, they emphasise them differently. The digital traces that people produce online can be read as a sensor of public attention \cite{galesic2021human}, and comparing that sensor against institutional reporting has become more pressing as agenda-setting and misinformation dynamics now play out across both kinds of platform at once \cite{doi:10.1126/science.aap9559,tsfati2020causes}. A natural way to make such a comparison is to model the topics present in each corpus and then compare the resulting topic distributions. This turns out to be harder than it first appears, and the difficulty is methodological rather than conceptual.

The obstacle is comparability. Topic models assign their own corpus-specific identifiers to the structure they discover, so a topic recovered from tweets and a topic recovered from news articles are not expressed in the same vocabulary of labels. Without a common reference layer, the two topic outputs cannot be placed side by side, and any comparison built on top of them inherits that incompatibility. The closest prior work, by Zhao et al.\ \cite{zhao2011comparing}, applied Latent Dirichlet Allocation independently to Twitter and The New York Times and compared the resulting distributions, finding partial overlap in broad themes but substantial differences in attention. That study established the value of cross-media topic comparison, but it rests on two properties that limit a controlled, repeatable comparison. It depends on a model family that tends to produce diffuse, overlapping topics on short, noisy social-media text, and on unsupervised topic discovery that yields a different topic space for each corpus, so the comparison space itself shifts with the data.

The first of these properties reflects a wider issue in topic modelling for short text. Classical bag-of-words models such as Latent Dirichlet Allocation \cite{blei2001latent} and Non-Negative Matrix Factorisation \cite{lee1999learning} rely on repeated word co-occurrence and tend to produce diffuse, overlapping topics on tweets \cite{hankar2025comprehensive}. Embedding-based alternatives address this by representing documents in dense semantic spaces. Top2Vec \cite{angelov2020top2vec} clusters joint document-word embeddings, whereas BERTopic \cite{grootendorst2022bertopic} combines sentence-transformer embeddings with dimensionality reduction and density-based clustering, and uses a class-based TF-IDF weighting to produce interpretable topic representations. Empirical comparisons on Twitter data report that BERTopic yields more coherent topics and fewer noisy clusters than LDA, NMF, or Top2Vec \cite{egger2022topic}. BERTopic also supports a guided variant, in which seed words shape topic boundaries, and a supervised variant for assignment to predefined classes \cite{GrootendorstGuidedBERTopic}; together these make consistent, like-for-like topic assignment across two corpora feasible.

This paper targets the second limitation, the corpus-specific topic space. Rather than letting each corpus define its own topics, it fixes a single shared topic space in advance, defined by the IPTC Media Topics taxonomy\footnote{\url{https://iptc.org/std/NewsCodes/guidelines/}}, the international standard used by news organisations to classify reporting, so the comparison space no longer shifts with the data. Both corpora are mapped into that taxonomy, anchoring the comparison in an externally validated reporting scheme rather than in labels introduced per corpus. Anchoring it in this way makes the framework reproducible and the topic labels directly interpretable. The topic model is guided BERTopic, which preserves unsupervised topic discovery while allowing topic boundaries to be shaped by the taxonomy, and the mapping is hierarchical: discovered topics are first scored against the fine-grained level-1 topics and then collapsed upward to the seventeen IPTC parent topics that serve as the reporting layer. The framework is developed and selected on a controlled New York Times 2011 corpus.

This paper makes the following contributions. First, it defines a shared topic space that maps heterogeneous corpora into the IPTC Media Topics taxonomy, so that social-media and news topic outputs become directly comparable. Second, it introduces a parent-enriched target construction together with a maximum-similarity collapse rule that moves from fine-grained level-1 scores to the seventeen reporting-level parent topics under conservative score and gap thresholds. Third, it sets out a development-corpus selection protocol, comprising a broad model screen, a strict finalist comparison, a target-construction ablation, and a threshold calibration, in which the guided family is preferred over a zero-shot baseline and the parent-enriched targets over simpler ones. 
Two questions organise the work: (1) which model family and mapping design produce adequate and defensible coverage of the shared topic space; and (2) whether the selected configuration depends on a single brittle threshold or remains usable across a range of assignment rules. 

\section{Methods}

The New York Times 2011 corpus serves as the development corpus on which the shared topic-space pipeline was designed and selected. 
%
%
News articles are long text pieces, editorially produced, and comparatively coherent, whereas tweets are shorter, noisier, and far more numerous. To pass both through one pipeline, the modelling inputs were reduced to a common text field for embedding and topic assignment, while publication timestamps were retained for later aggregation. The unit of analysis changes across the pipeline. At the document level, each article or tweet receives a mapped IPTC parent label or remains an outlier. At the daily level, topic shares form time series. At the monthly level, mapped document counts become seventeen-topic share vectors. The method is therefore built around a hierarchy of representations rather than a single fixed one, with source harmonisation as the central design requirement.

\begin{figure}[ht]
    \centering
    \includegraphics[width=\textwidth]{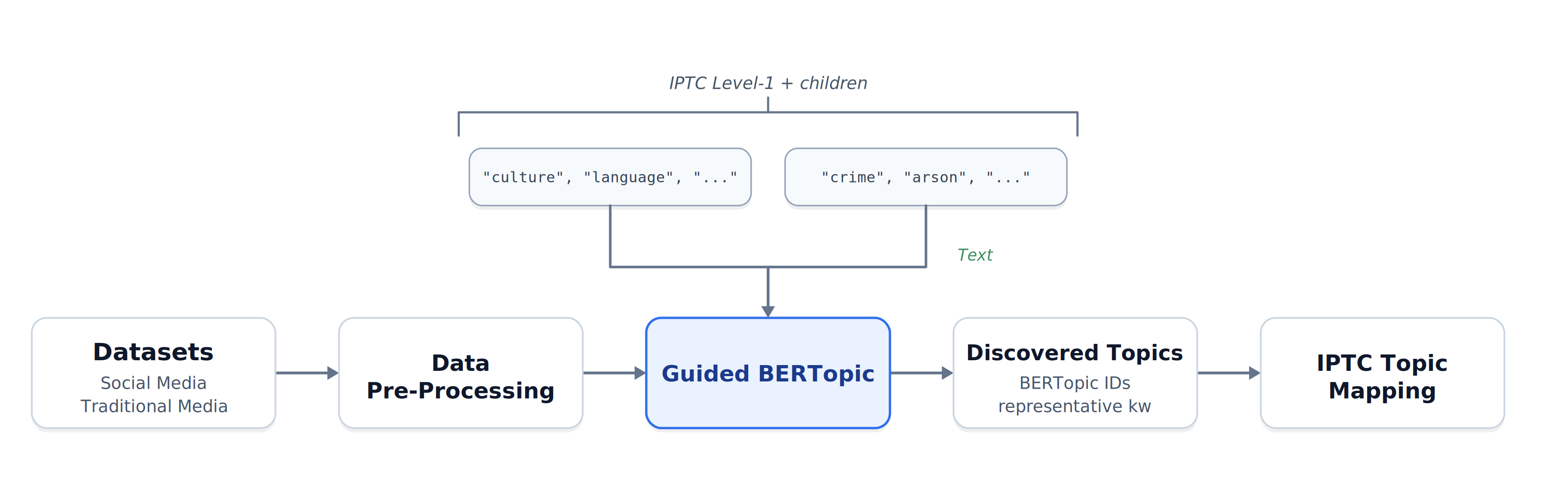}
\caption{Overview of the shared topic-space workflow. The corpora are pre-processed, modelled with guided BERTopic, and mapped into IPTC labels to place both sources in the common topic space.}    \label{fig:method_pipeline}
\end{figure}

The comparison space is defined by the IPTC Media Topics hierarchy\footnote{\url{https://iptc.org/std/NewsCodes/guidelines/}}, and results are reported at the level of the seventeen IPTC parent topics. This level is broad enough to support stable cross-source comparison while remaining interpretable as a reporting taxonomy. Assignment is not made directly to those seventeen labels. Instead, discovered topics are first scored against the ninety-four IPTC level-1 topics and then collapsed upward to their parent topics. Scoring at the finer level preserves more semantic detail and reduces the arbitrariness of direct top-level matching. Guided BERTopic provides the base topic model \cite{grootendorst2022bertopic,GrootendorstGuidedBERTopic}, preferred because it preserves unsupervised topic discovery while allowing topic boundaries to be shaped by the taxonomy. For each level-1 topic, the seed bag contains the level-1 label and its level-2 child terms. In the final mapping layer, the parent label is also included directly in the target centroid, so target construction is parent-enriched rather than child-only, matching the goal of reliable assignment to the seventeen parent topics. Figure~\ref{fig:method_pipeline} summarises the workflow.

Once BERTopic has discovered a set of topics, each discovered topic $i$ is represented by a weighted centroid $v_i$ built from its top keyword set. Each IPTC level-1 target $j$ is represented by a centroid $c_j$ built from the parent term, the level-1 label, and the associated level-2 child terms, with higher weight assigned to the parent and level-1 labels than to individual child terms. The topic-to-target similarity is the cosine similarity,
\begin{equation}
s_{ij} = \cos(v_i, c_j).
\end{equation}
If parent topic $p$ contains the set of level-1 children $C(p)$, the parent-level score is
\begin{equation}
S_{ip} = \max_{j \in C(p)} s_{ij}.
\end{equation}
This \texttt{parent\_max} rule is the core collapse mechanism: it preserves the strongest child-level evidence while keeping the reporting layer fixed at the seventeen IPTC parents.

The previous paragraphs describe the proposed mapping framework. The remainder of this section describes how the final configuration of this framework was selected on the NYT 2011 development corpus before being deployed unchanged to the evaluation corpora.

Method selection on NYT 2011 followed a narrowing sequence rather than a single comparison. An initial broad screen compared guided and zero-shot variants across alternative document representations and topic-representation models, using the mapped-document rate as a first-pass viability filter. This screen was not intended to identify a final model directly; its purpose was to eliminate model families that failed to produce adequate mapped coverage or collapsed to too few usable parent labels. A focused refinement step then evaluated how discovered topics should be mapped into IPTC, comparing weighted keyword centroids, representative-document centroids, hybrid topic vectors, direct top-level assignment, hierarchical level-1-to-parent mapping, lexical overlap routes, and alternative collapse rules. Refinement and the later finalist stage considered the mapped rate alongside parent-label recovery, similarity strength, gap strength, and parent consistency, so that coverage was never the sole criterion. The strongest candidates were rerun under stricter post-hoc thresholds, and a final threshold sweep calibrated the assignment rule for the retained guided route. Two decisions are central, and both were made on the development corpus rather than on downstream alignment outcomes: the framework uses the guided BERTopic family rather than the zero-shot alternative, and it uses hierarchical IPTC mapping rather than direct parent-level matching.

Target construction was treated as a separate mapping problem. On the fixed guided run, a post-hoc ablation compared three target constructions: \texttt{level1}, \texttt{level1 + level2}, and \texttt{parent + level1 + level2}. The parent-enriched variant was retained because it improved mapped coverage and parent consistency on the development corpus, a choice made on mapping quality rather than on whether it made the two sources appear closer. 

The final configuration uses guided BERTopic with level-1 seed bags enriched by level-2 child terms, sentence-transformer embeddings from \texttt{all-MiniLM-L6-v2}\footnote{\url{https://huggingface.co/sentence-transformers/all-MiniLM-L6-v2}}, UMAP~\cite{mcinnes2018umap} with \texttt{n\_neighbors = 15}, \texttt{n\_components = 5}, and \texttt{min\_dist = 0.0}, cosine distance, Maximal Marginal Relevance (MMR) topic representation~\cite{carbonell1998use} with diversity \texttt{0.35}, and HDBSCAN~\cite{mcinnes2017hdbscan} clustering. For the final annual news model, \texttt{min\_cluster\_size = 300} was retained. Each discovered topic was represented by a weighted centroid over its top ten keywords and mapped through the parent-enriched target space using \texttt{parent\_max}. A parent assignment was retained only when the best score and the gap to the runner-up both exceeded fixed thresholds:
\begin{equation}
S_{i(1)} \geq 0.60, \qquad S_{i(1)} - S_{i(2)} \geq 0.02.
\end{equation}
These thresholds removed ambiguous assignments without reducing mapped coverage to unusable levels.

\section{Results}

\begin{figure}[ht]
    \centering
    \includegraphics[width=.92\textwidth]{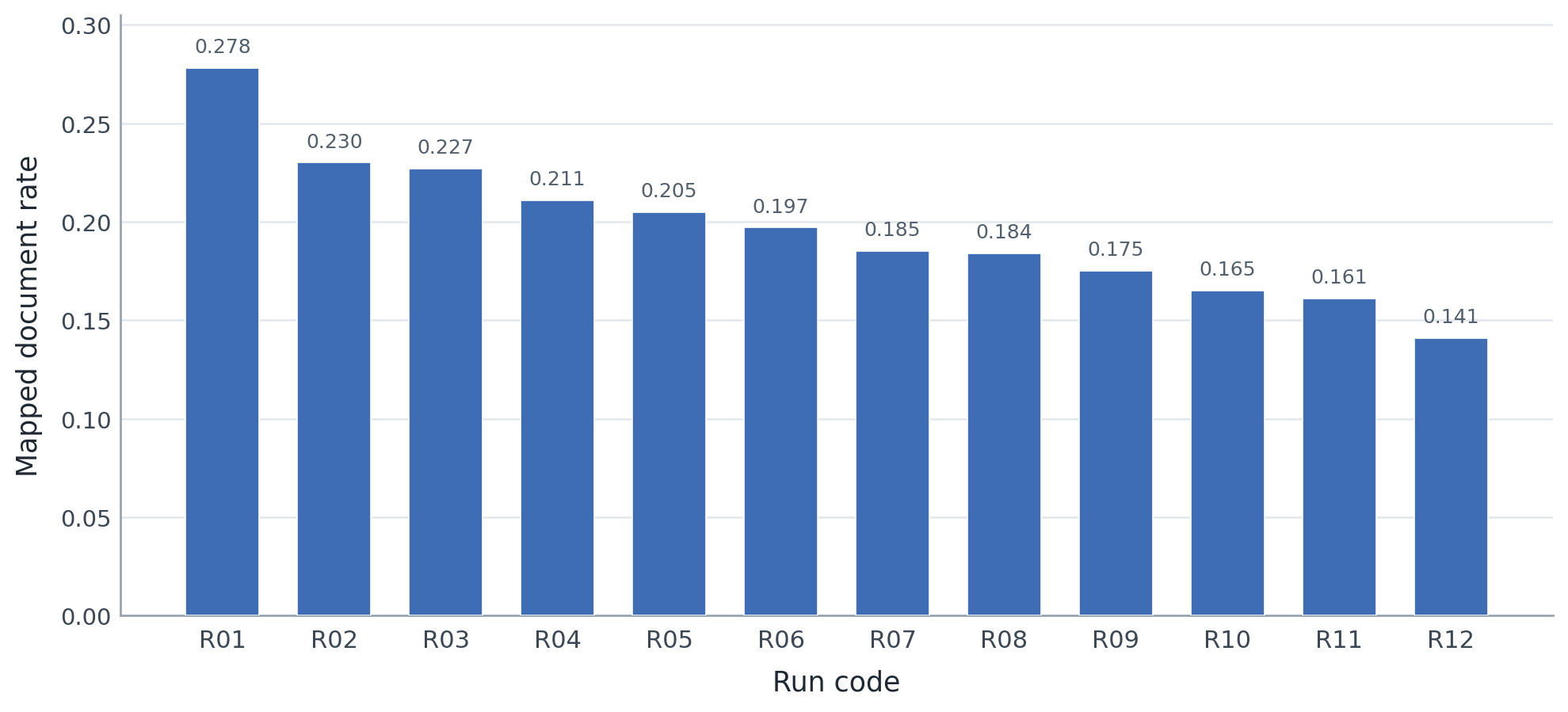}
\caption{Broad BERTopic screen on the NYT 2011 development corpus. Each run combines a modelling family, a document representation, and a clustering specification, ordered by the mapped-document rate as an initial viability filter before more detailed refinement. 
}\label{fig:broad_screen}
\end{figure}

Model selection on the development corpus began with a broad screen, shown in Figure~\ref{fig:broad_screen}, the first and coarsest stage of the selection sequence, intended to rule out unsuitable model families before the more detailed comparisons that follow. Each run combined a modelling family, a document representation, and a clustering specification, and the mapped-document rate ordered the runs as an initial viability filter. Position in this ranking reflects mapped coverage alone, so the highest-ranked run is not necessarily the selected model; the final configuration was fixed only by the later, more discriminating stages. The screen showed that guided configurations were consistently more suitable than the zero-shot baselines once the objective was stable hierarchical mapping rather than unrestricted topic discovery alone. This eliminated the weaker families and concentrated the subsequent refinement on the stronger guided candidates and on the mapping layer itself. Because the screen ranked runs only by this first-pass filter, it defined the usable region within which those comparisons were then carried out.

\begin{figure}[ht]
    \centering
    \includegraphics[width=\textwidth]{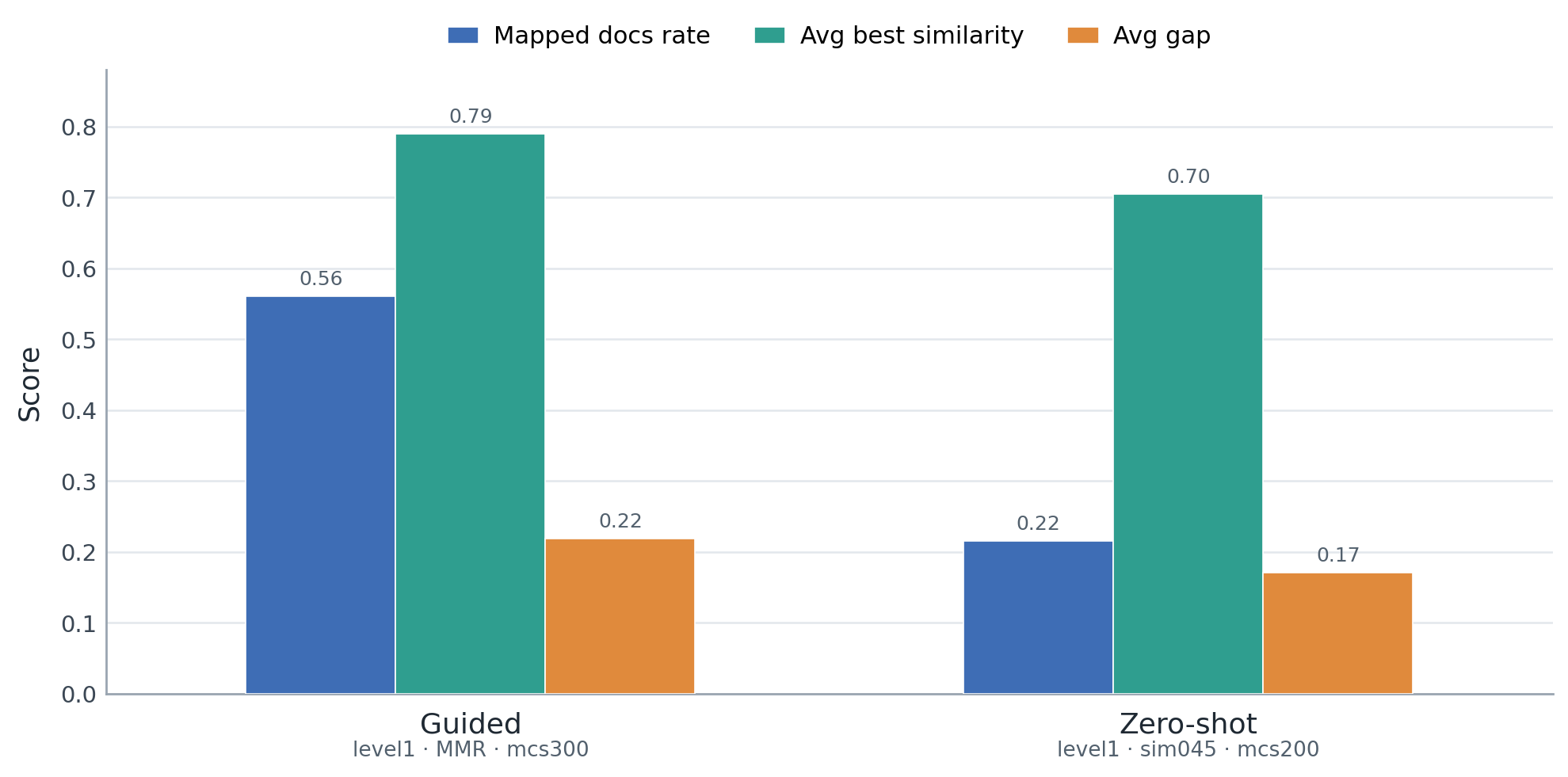}
\caption{Strict finalist comparison on the development corpus. The guided level-1 MMR model retained substantially stronger mapped coverage than the zero-shot benchmark under stricter assignment thresholds, which is why the guided family was carried forward into the annual pipeline.}
\label{fig:finalists_strict}
\end{figure}

The finalist stage then compared the retained guided route against a zero-shot benchmark under stricter assignment thresholds, shown in Figure~\ref{fig:finalists_strict}. Three quantities were reported together: the mapped-document rate, the average best similarity of mapped topics to their assigned target, and the average gap between the best and runner-up targets. The guided level-1 MMR model retained substantially stronger mapped coverage than the zero-shot benchmark while maintaining comparable average best similarity and a usable gap to the runner-up. The combination of higher coverage and adequate similarity under the stricter rule was the reason the guided family was carried forward into the annual pipeline, since a model that maps far fewer documents would leave the downstream topic-share series sparse and unstable.

\begin{figure}[ht]
    \centering
    \begin{minipage}[t]{0.48\textwidth}
        \centering
        \includegraphics[width=\textwidth]{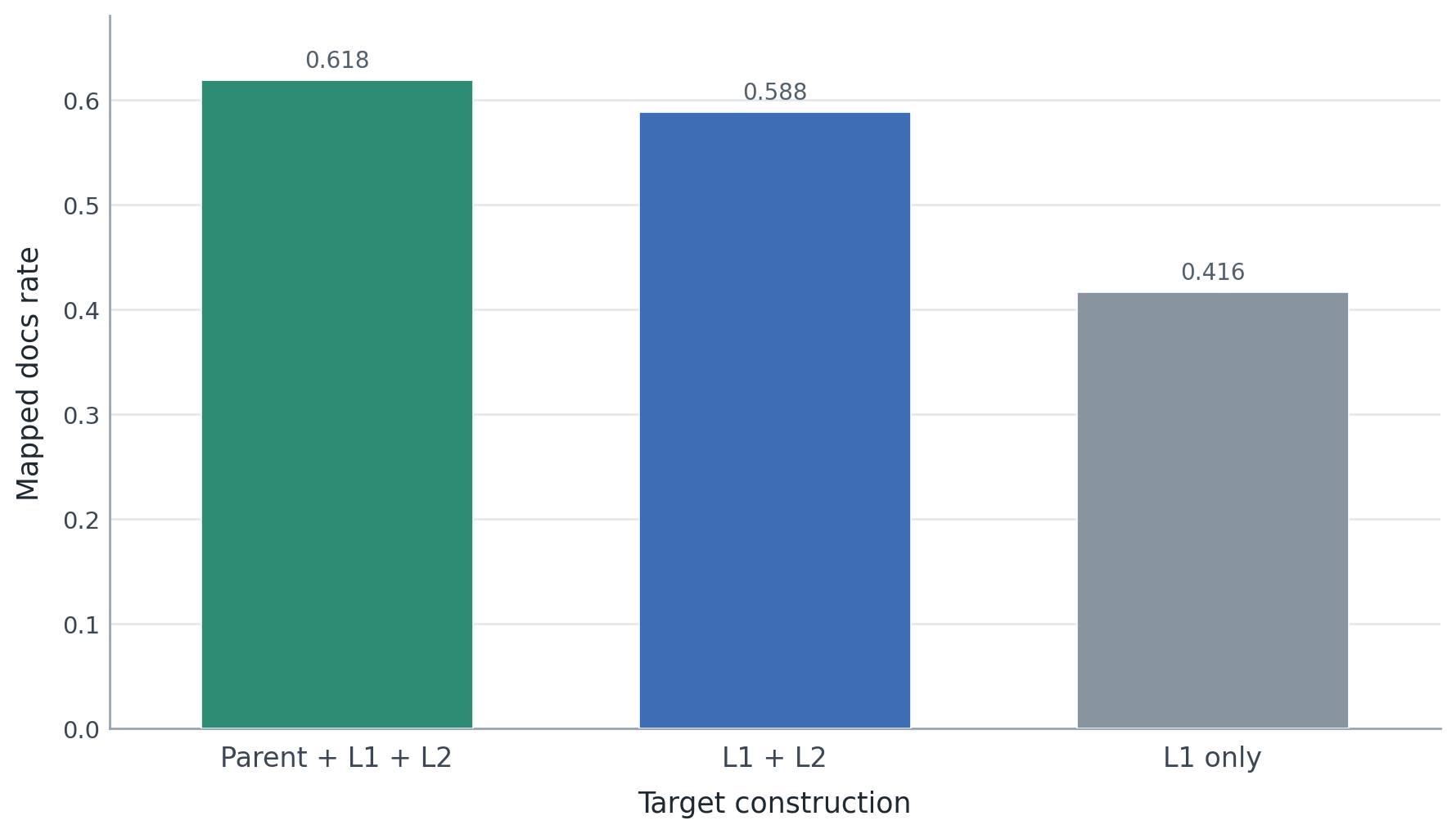}
    \end{minipage}\hfill
    \begin{minipage}[t]{0.48\textwidth}
        \centering
        \includegraphics[width=\textwidth]{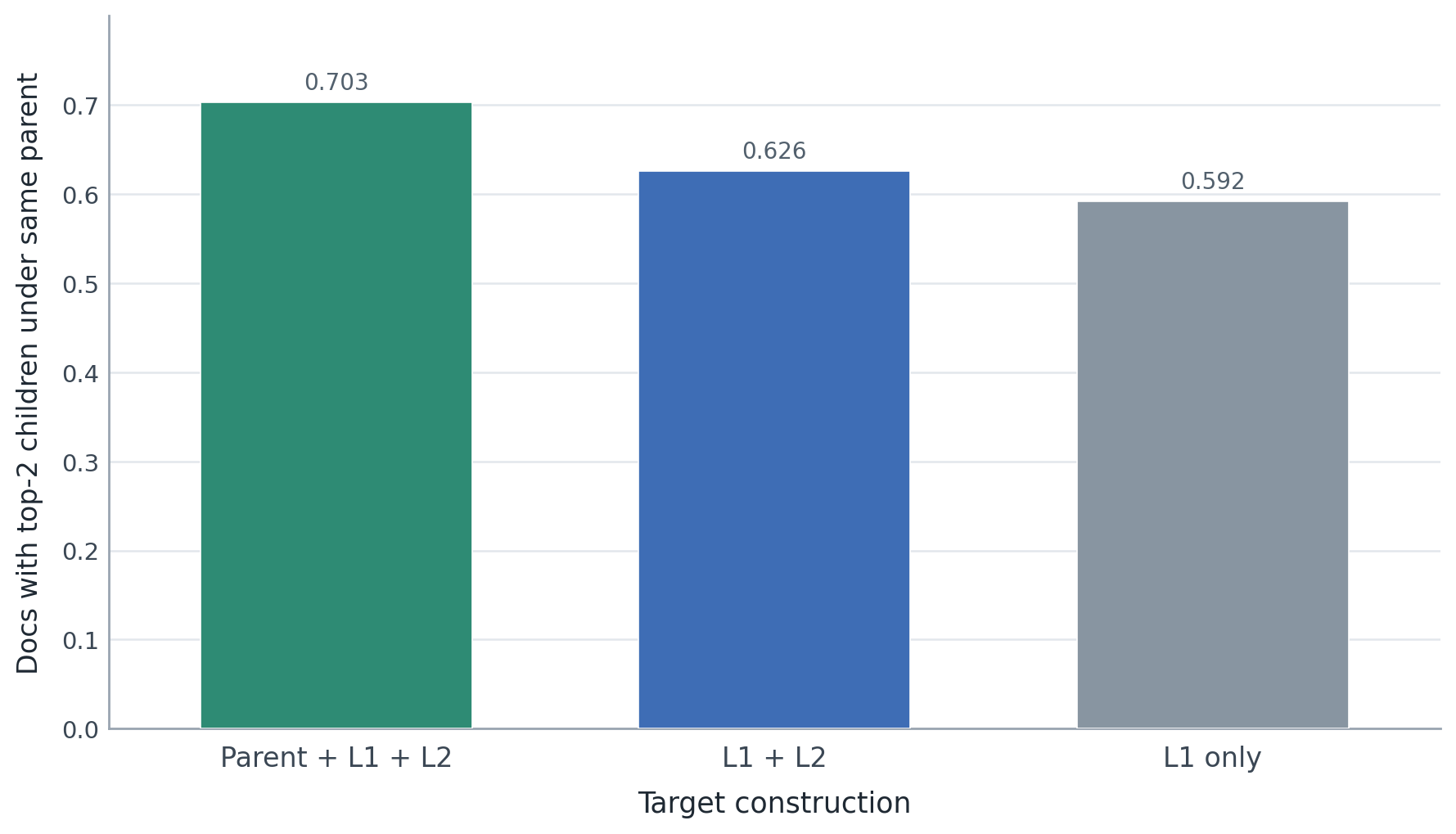}
    \end{minipage}
\caption{Target-space ablation on the fixed guided run. The parent-enriched construction improved mapped coverage (left) and strengthened parent-level top-two agreement (right) relative to the simpler target variants, which is why the final analysis uses the \texttt{parent + level1 + level2} target space.}
\label{fig:target_ablation}
\end{figure}

With the guided family fixed, the target-construction ablation compared the three target spaces on coverage and on parent consistency, shown in Figure~\ref{fig:target_ablation}. The parent-enriched \texttt{parent + level1 + level2} construction produced the highest mapped-document rate of the three. It also produced the strongest parent-level top-two agreement, measured as the share of documents whose top-two children fall under the same parent, which rose from $0.592$ for \texttt{level1} to $0.626$ for \texttt{level1 + level2} and $0.703$ for the parent-enriched construction. Higher coverage and stronger parent consistency pointed in the same direction, and together they justified the parent-enriched target space used in the annual analysis. The two panels also showed that the simpler \texttt{level1} construction, while interpretable, left a larger share of documents either unmapped or split inconsistently across parents.

\begin{figure}[ht]
    \centering
    \includegraphics[width=.88\textwidth]{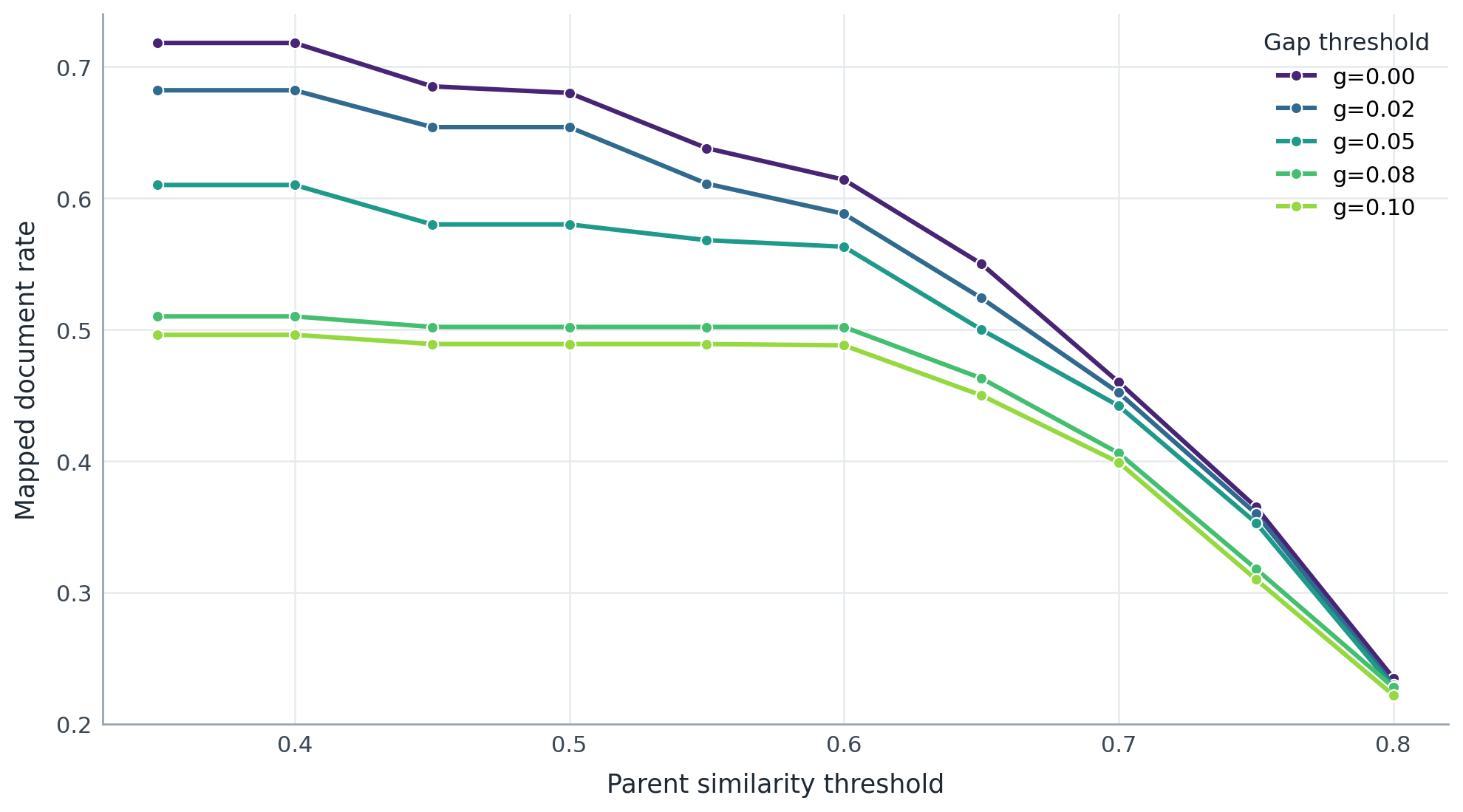}
\caption{Threshold sweep for the retained guided route. The figure shows how mapped-document coverage changes as the parent-assignment threshold is tightened, supporting the final choice of a conservative but still usable mapping rule.}
\label{fig:threshold_sweep}
\end{figure}

The final step was a threshold sweep on the retained guided route, shown in Figure~\ref{fig:threshold_sweep}. As the parent-assignment threshold was tightened, and across a range of gap thresholds, mapped coverage declined gradually rather than collapsing at any single point. A gradual decline matters because it indicates that the selected assignment rule sits on a smooth part of the trade-off between ambiguity control and usable coverage, so the configuration does not depend on one brittle threshold setting. The sweep therefore functioned as a sensitivity check on the final rule rather than as a separate model-selection stage.

Taken together, the development-corpus results selected a single configuration: guided BERTopic with parent-enriched hierarchical IPTC mapping, a maximum-similarity parent collapse, and conservative score and gap thresholds. The four stages are complementary rather than redundant. The broad screen fixed the viable model family, the finalist comparison confirmed that the guided route survives stricter assignment, the target ablation fixed the construction of the mapping targets, and the threshold sweep confirmed that the final rule is not brittle. When this fixed configuration was deployed to the annual evaluation corpora, each annual run was fit once and aggregated into comparable monthly and daily topic-share series for both sources within one shared seventeen-topic space. The framework therefore delivers the time-indexed, like-for-like topic-share representations that downstream alignment analysis requires.

\section{Discussion}
The central outcome of this work is a defensible shared topic space. The mapping decisions, namely the guided rather than zero-shot family, hierarchical rather than direct parent matching, parent-enriched rather than child-only targets, and a maximum-similarity collapse with conservative thresholds, were all made on a development corpus and on mapping-quality criteria rather than on whether they made the two sources look more similar. This separation matters for a comparison study, because tuning the comparison space on the same pair that is later compared would risk manufacturing alignment. Because both corpora are mapped into the same externally defined reporting taxonomy, their topic outputs become directly comparable, which is the precondition for any subsequent cross-source comparison. The methodological contribution is therefore the construction of the comparison space itself, in a form that is reproducible and interpretable.

More broadly, the choices made here instantiate a general template for comparable topic modelling, of which this study demonstrates a single instance. The essential requirement is not the IPTC taxonomy in particular, nor guided BERTopic, nor the Twitter and news pairing, but prior agreement on a shared set of topics that defines the comparison space. Given such a reference set, heterogeneous corpora can each be modelled by whatever discovery method best suits their text, their discovered topics scored against the agreed categories, and the result collapsed to a common reporting level for like-for-like comparison. The reference taxonomy, the topic model, and the collapse rule are interchangeable components, so the framework extends to any setting in which the sources being compared can agree, in advance, on the categories that form the basis of comparison. This decoupling of how each corpus is modelled from what the corpora are compared on is what makes the comparison reproducible and portable across domains, beyond the social-media and news case studied here.

Several limitations stem from the design choices. The shared topic space involves a trade-off. Using the IPTC taxonomy improves interpretability and cross-source comparability, but it compresses platform-specific discourse into a relatively broad reporting scheme. This is appropriate for comparing topic attention, though less suited to capturing subtle differences in framing, stance, or rhetorical style. 
The Twitter corpus is a large stratified annual sample rather than a complete archive of platform activity, and the news side is a single mainstream outlet, so the framework is demonstrated as a controlled case rather than validated across many outlets. 

Several extensions follow naturally. The most immediate is to apply the framework to additional mainstream outlets and to multiple years, allowing the stability of the recovered topic space to be tested across different institutional contexts and event cycles. Relatedly, broader validation of the annual social-media model across further completed runs would strengthen confidence in the deployed configuration. A second direction is to enrich the shared layer itself, for example by retaining finer IPTC levels for selected analyses, so that the space supports comparisons of framing and emphasis in addition to topic attention.

\section{Conclusion}
This paper presented a reproducible framework for constructing a shared topic space based on the IPTC Media Topics taxonomy. Using the NYT 2011 development corpus, we showed that (i) guided BERTopic consistently outperformed a zero-shot alternative for hierarchical topic mapping, (ii) parent-enriched target representations improved both mapped coverage and parent consistency, and (iii) the selected mapping remained robust across a range of assignment thresholds. Together, these results establish a principled methodology for building externally anchored topic spaces that enable consistent cross-corpus topic comparison. Future work will evaluate the framework across additional corpora, domains and taxonomies.
\bibliographystyle{splncs04}
\bibliography{main}

\end{document}